\documentclass[preprint,12pt,a4paper,authoryear]{elsarticle}
\usepackage[utf8]{inputenc}

\usepackage{hyperref}
\hypersetup{
    bookmarksopen=true,
    bookmarksnumbered=true,
    breaklinks=true,
    hidelinks=true,
}

\usepackage[acronym,nomain,nopostdot]{glossaries}
\newacronym{ac}{AC}{alternating current}
\newacronym{acer}{ACER}{European Agency for the Cooperation of Energy Regulators}
\newacronym{admm}{ADMM}{Alternating Direction Method of Multipliers}
\newacronym{anova}{ANOVA}{analysis of variance}
\newacronym[firstplural={available transfer capacities (ATCs)},plural=ATCs]{atc}{ATC}{available transfer capacity}
\newacronym{bnetza}{BNetzA}{German Federal Network Agency}
\newacronym{ca}{CA}{capacity allocation}
\newacronym{cbco}{CBCO}{critical branch under critical outage}
\newacronym{cb}{CB}{critical branch}
\newacronym{co}{CO}{critical outage}
\newacronym{cacm}{CACM}{capacity allocation and congestion management}
\newacronym{cee}{CEE}{Central Eastern Europe}
\newacronym{cm}{CM}{congestion management}
\newacronym{cne}{CNE}{critical network element}
\newacronym[firstplural={critical network elements and contingencies (CNECs)},plural=CNECs]{cnec}{CNEC}{critical network element and contingency}
\newacronym{co2}{CO\textsubscript{2}}{carbon dioxide}
\newacronym{core}{CORE}{CORE region}
\newacronym[firstplural=Central Western European (CWE)]{cwe}{CWE}{Central Western Europe}
\newacronym{d2cf}{D2CF}{2-days ahead congestion forecast}
\newacronym{da}{DA}{day-ahead}
\newacronym{dcpf}{DCPF}{linear power flow}
\newacronym{dsk}{DSK}{demand shift key}
\newacronym{der}{DER}{distributed energy resource}
\newacronym{ec}{EC}{European Commission}
\newacronym{eex}{EEX}{European Energy Exchange}
\newacronym{euphemia}{EUPHEMIA}{Pan-European Hybrid Electricity Market Integration Algorithm}
\newacronym{fav}{FAV}{final adjustment value}
\newacronym{fba}{FBA}{flow-based allocation}
\newacronym{fbmc}{FBMC}{flow-based market coupling}
\newacronym{flh}{FLH}{full load hour}
\newacronym{ffe}{FfE}{Forschungsstelle für Energie}
\newacronym{frm}{FRM}{flow reliability margin}
\newacronym{gsk}{GSK}{generation shift key}
\newacronym{hev}{HEV}{hybrid electric vehicle}
\newacronym{hvdc}{HVDC}{high-voltage direct-current}
\newacronym{iem}{IEM}{internal energy market}
\newacronym{jump}{JuMP}{Julia for Mathematical Programming}
\newacronym{jao}{JAO}{Joint Allocation Office}
\newacronym{lp}{LP}{linear program}
\newacronym{lodf}{LODF}{load outage distribution factor}
\newacronym{lsk}{LSK}{load shift key}
\newacronym{isk}{ISK}{injection shift key}
\newacronym{mio}{mio.}{million}
\newacronym{minram}{minRAM}{minimum remaining available margin}
\newacronym{ndp}{NDP}{network development plan}
\newacronym[firstplural={net transfer capacities (NTCs)}, plural=NTCs]{ntc}{NTC}{net transfer capacity}
\newacronym{nuts}{NUTS}{Nomenclature of Territorial Units for Statistics}
\newacronym{opsd}{OPSD}{Open Power System Data}
\newacronym{or}{OR}{Operations Research}
\newacronym{opf}{OPF}{optimal power flow}
\newacronym{otc}{OTC}{over-the-counter}
\newacronym{pcr}{PCR}{Price Coupling of Regions}
\newacronym[sort=pomato]{pomato}{\begin{small}POMATO\end{small}}{Power Market Tool}
\newacronym{psp}{PSP}{pumped-storage plants}
\newacronym{pv}{PV}{photovoltaik}
\newacronym{ptdf}{PTDF}{power transmission distribution factor}
\newacronym{px}{PX}{power exchange}
\newacronym{ram}{RAM}{remaining available margin}
\newacronym{res}{RES}{renewable energy sources}
\newacronym{scopf}{SCOPF}{security-constrained optimal power flow}
\newacronym{tso}{TSO}{transmission system operator}
\newacronym{tyndp}{TYNDP}{Ten-Year Network Development Plan}

\glsdisablehyper

\usepackage{mathtools}
\usepackage{amsmath, amssymb, amsthm, bm}  
\usepackage{float}
\usepackage{multirow}
\usepackage{booktabs}
\usepackage{tabularx}

\usepackage[smaller]{acronym}
\usepackage{svg}
\usepackage{xcolor}
\usepackage{url}

\usepackage[noabbrev,capitalize]{cleveref}

\crefname{equation}{}{}
\Crefname{equation}{}{}

\usepackage{currency}
    \DefineCurrency{EUR}{name={euro}, plural={euros}, symbol={\euro}, iso={EUR}, kind=iso}

\usepackage{listings}

\definecolor{pyred}{RGB}{186, 33, 33}
\definecolor{pygreen}{RGB}{0, 128, 0}
\definecolor{pyblue}{RGB}{0, 0, 255}
\definecolor{pyteal}{RGB}{64, 128, 128}

\lstdefinestyle{Python}{
    language        = Python,
    basicstyle      = \small\ttfamily,
    keywordstyle    = \color{pygreen},
    keywordstyle    = [2] \color{pygreen}, 
    stringstyle     = \color{pyred},
    commentstyle    = \color{pyteal}\ttfamily
}
\lstdefinestyle{JSON}{
    language        = Python,
    basicstyle      = \small\ttfamily,
    keywordstyle    = \color{pygreen},
    stringstyle     = \color{pyred},
}

\newcommand{\set}[1]{\mathcal{#1}} 
\DeclareMathOperator{\PTDF}{PTDF}

\DeclareMathOperator{\EX}{EX}

\theoremstyle{definition} 

\theoremstyle{plain} 
\theoremstyle{remark} 

\definecolor{bg}{rgb}{0.95,0.95,0.95}

\begin{document}
\sloppy 

\title{Evaluating Policy Implications on the Restrictiveness of Flow-based Market Coupling with High Shares of Intermittent Generation: A Case Study for Central Western Europe}

\author[TUWIP]{Richard Weinhold\corref{mycorrespondingauthor}}
\ead{riw@wip.tu-berlin.de}
\cortext[mycorrespondingauthor]{Corresponding author}
\address[TUWIP]{Fakultät VII Wirtschaft und Management, TU Berlin, 10623 Berlin, Germany.}

\begin{frontmatter}
\begin{abstract}
The current stage in the evolution of the European internal energy market for electricity is defined by the transformation towards a renewable energy system. The Clean Energy Package aims to ensure that methods for capacity allocation and congestion management, that are at the center of the European internal market for electricity, align with this transformation. 

Flow-based market coupling, the preferred method for capacity allocation, is first and foremost a formal process to allocate exchange capacities to the markets. However, the process also allows for many considerations of the involved parties that impact the resulting capacities. As part of the Clear Energy Package, the regulatory body enacted their ambition to increase exchange capacities by enforcing transmission system operators to allocate a minimum margin of physical line capacity with the goal of providing a higher level of competition and better integration of renewable energy sources. This study investigates this and other policy relevant consideration of flow-based market coupling.

The model results quantify the trade-off between permissive capacity allocation and increased congestion management. For high shares of intermittent renewable generation, less constrained exchange capacities are favorable, however also highlight the importance of the markets ability to integrate high shares of intermittent generation.
\end{abstract}


\end{frontmatter}

\section{Introduction}
\glsresetall

Europe's commitment to become climate neutral by 2050 \citep{directorategeneralforenergy_clean_2019} draws a clear path towards a fully decarbonized electricity sector. This transformation is characterized by a large increase of intermittent renewable generation as well as decommissioning of conventional and nuclear generation capacities. 
The current understanding of the European \gls{iem} for electricity aims to efficiently achieve climate targets while providing generators and consumers with non-discriminatory market access and ensuring affordable and secure provision of electricity \citep[Art. 1]{europeancommission_commission_2019}. Its central mechanism of capacity allocation and congestion management aims to efficiently use transmission infrastructure, ensures operational security and transparency \citep{europeancommission_commission_2015}.
Capacity allocation summarizes methods and regulations that dimension electricity trading volumes that market participants can use based on the physical transmission capacity and operational considerations. Congestion management describes actions and protocols taken by the responsible \glsreset{tso}\gls{tso} if network congestion occurs \citep{europeancommission_directive_1997}.  

Previously, capacity allocation was implemented using static, bilateral \glspl{ntc} which are based on non-public network models and assessments of historic network loads \citep{etso_procedures_2001}. \glspl{ntc} do not account for restricting transmission assets within market zones, resulting in potentially too conservative capacity allocation and increased congestion management \citep{amprion_cwe_2011}. 
From 2015 \glsreset{fbmc}\gls{fbmc} replaced \gls{ntc} as the preferred method for capacity allocation and is used in the \glspl{cwe} region\footnote{The \gls{cwe} region consists of Belgium, France, Germany, Luxembourg and the Netherlands.}. Its main advantages are increased transparency from a clear methodology that describes capacity allocation and, more importantly, the fact that capacity is allocated towards the net-position of each bidding-zone based on individual network elements rather than bilaterally. Thus \gls{fbmc} ultimately better aligns with the goal to utilize the network infrastructure more efficiently and accommodate the transformation towards a decarbonized electricity system. 

Since its inauguration, \gls{fbmc} has proven to be the more efficient capacity allocation compared to \glspl{ntc} while providing at least the same level of security \citep{rte_cwe_2015}. However, the regulation seeks to achieve high levels of price convergence between market areas leading to overall lower prices, unrestricted access to the internal European electricity market and thereby successful integration of renewable generation \citep{acer_annual_2020}. As a result, Regulation \citep{europeancommission_commission_2019} makes clear that capacity allocation has priority over congestion management and that \glspl{tso} should not restrict commercial exchange to solve internal congestion, as previously observed by \citet{acer_decision_2016}, and explicitly requires that at least 70\% of physical capacity is allocated to the market. This directly alters the method to derive trading volumes and makes clear that the formal process of deriving trading volumes using the flow-based methodology, which originally made the appearance to be a purely formal process, leaves room to implement policy decisions. While the canon of academic literature explains and depicts the formal side of the process and explores the effects of different parametrizations very well, policy decisions that define which outcome aligns with political targets are rarely discussed or numerically modeled. 

Capacity allocation within \gls{fbmc} is defined by the two objectives of market integration, meaning the provision of commercial exchange capacity, and secure operation whose outcome results in very different trading volumes. This paper aims to highlight this presumed trade-off and investigate how the policy decision to prioritize capacity allocation over congestion management affects the efficiency of \gls{fbmc} and its ability to accommodate the transformation in the European electricity system. 

\section{Background on FBMC and Literature Review}\label{sec:cwe-modeling_fbmc}

\begin{figure}[H]
  \centering
    \includegraphics[width=0.95\textwidth]{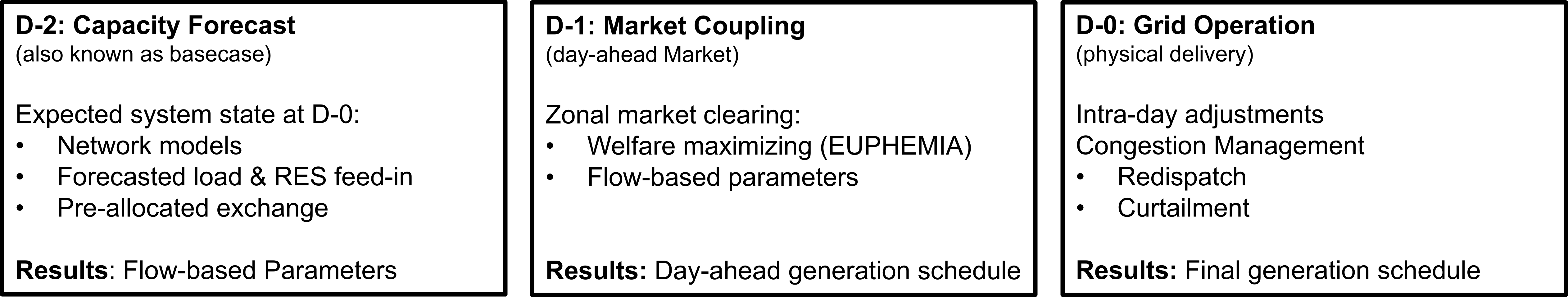}
  \caption{Flow-based market coupling process. Based on \citet{amprion_amprion_2019}}
  \label{fig:cwe-fb_process}
\end{figure}

\gls{fbmc} is a multistage process coordinated by the \glspl{tso} which aims to allocate commercial exchange capacities to the markets. Specifically, this three step process, as depicted in Figure~\ref{fig:cwe-fb_process}, consists of a D-2 capacity forecast, also called basecase, which represents the best estimate of the system state at delivery \citep{50hertz_documentation_2020}. The basecase is a result from forecasts on load and \gls{res} feed-in and already allocated capacities e.g. long term nominations in conjunction with network models. Based on this forecast so called flow-based parameters are calculated and used to constrain the commercial exchange in the day-ahead market coupling stage. In the European \gls{iem} the central market clearing algorithm \gls{euphemia} matches demand and supply bids subject to the flow-based parameters and maximizing welfare. Lastly, D-0 stage consists of intraday adjustments and congestion management, meaning the physical delivery. 

Of particular interest are the so called flow-based parameters, which constitute the capacity allocation in \gls{fbmc} and are generated, exchanged and published following specific rules requested by regulation \citep{europeancommission_commission_2015}. Details on how the parameters are obtained is laid out in the documentation of the flow-based process \citep{50hertz_documentation_2020}. 
Their composition is defined by the remaining available commercial exchange capacity, denoted as \glsreset{ram}\gls{ram}, in the expected market outcome (basecase) which are allocated towards the markets, i.e. to changes in net-position $np$ between the forecasted basecase ($bc$) and day-ahead market coupling stage ($da$). Net-position changes are mapped to power flows in line elements by a so-called zonal \glsreset{ptdf}\gls{ptdf} matrix and bounded by line capacities $\overline{f}$ 

Equations \eqref{eq:cwe-fb_parameters} formalise this intuition, i.e. remaining capacities on network elements are available changes in net-positions between day-ahead and basecase. Reformulation yields Equation~\eqref{eq:cwe-fb_parameters_3} that defines the flow-based parameters in terms of net-positions in the day-ahead market stage, the zonal \gls{ptdf} matrix and \glspl{ram} and align with the formulation in \cite[p.55]{50hertz_documentation_2020}.

\begin{subequations}\label{eq:cwe-fb_parameters}
    \begin{align}
        \PTDF^z (np^{da} - np^{bc}) &\leq  \overline{f}  - f^{bc} \label{eq:cwe-fb_parameters_1} \\
        \PTDF^z np^{da} &\leq \overline{f} - f^{bc} + \PTDF^z np^{bc} \label{eq:cwe-fb_parameters_2} \\
        \PTDF^z np^{da} &\leq \overline{f} - f^{ref} = RAM \label{eq:cwe-fb_parameters_3} 
    \end{align}
\end{subequations}

A \gls{ptdf} matrix expresses the physical relation between nodal power injections and power flows in the transmission network in a linear manner \citep{weinhold_fast_2020}. The \gls{ptdf} is composed of a selection of network elements and contingencies and is thereby part of the parametrization. Commonly, \glsreset{cne}\glspl{cne} are selected based on their importance to commercial exchange. For each \gls{cne} a set of contingencies (C) can be considered if its outage significant impacts the respective \gls{cne}. The resulting set of \glsreset{cnec}\glspl{cnec} comprises the \gls{ptdf} matrix. Given that the net-position delta between basecase and day-ahead is small, the set of generators that will serve this delta by ``shifting'' their output can be anticipated. The resulting participation factor is called \glsreset{gsk}\gls{gsk} and is used as a mapping to transform the nodal $\PTDF^n$ to zonal $\PTDF^z =  \PTDF^n \cdot GSK$.

The resulting flow-based parameters define a feasible region for net-positions within the day-ahead market stage.

\begin{align}
    \set{F}^z(\PTDF^z, RAM) &= \{ x : \PTDF^z x \leq RAM\} &&\forall t \in \set{T}\label{eq:cwe-F_zonal}
\end{align}

The resulting feasible region \eqref{eq:cwe-F_zonal} has to be at least non-empty, which is not necessarily given, and either requires specific parametrization of the basecase, e.g. enforcing margins on \gls{cne} or processing of the \gls{ptdf} and \gls{ram}. This can be done by selecting specific networks elements that compose the \gls{ptdf}, enforce minimum \gls{ram} values. 

Therefore, the resulting flow-based parameters do not only reflect formal context but also methods to account for uncertainty or imperfections, e.g. from the zonal projections which allow or explicitly are used for the consideration of policy decisions towards the restrictiveness of commercial exchange. 
Generally, differences in parametrization can be explained by three reasons:
\begin{enumerate}
    \item To ensure secure operation and reduce congestion management. This could be done through the addition of
    security margins on \gls{ram} or by selecting a larger set of \glspl{cnec}, including internal network elements.
    \item To enlarge the trading domain and provide increased capacity to the market. This is more in line with regulation, and aims to achieve a higher level of price convergence. This is done through minimum values for RAM, formally knows as the \glsreset{minram}\gls{minram} criterion.
    \item To be more accurate, for example reducing inaccuracies of the zonal projection via \glspl{gsk}, by more precisely derived security margins based on historic data.
\end{enumerate}

Academic publications on modeling \gls{fbmc} and its parametrizations are still scarce. Early publications such as \citet{vandenbergh_flowbased_2016} and \citet{boury_methods_2015} formally describe the process and are extended recently by \citet{schonheit_fundamental_2021, felten_flowbased_2021, byers_modeling_2020} that discuss input parameters and their effects, illustrated by stylized examples. Generally, most academic publications adhere to a similar methodology that is depicted in Figure \ref{fig:cwe-fb_process} and utilize an economic dispatch problem for each of the three steps.

Applications based around the status-quo in the \gls{cwe} region exist and usually compare a result metric based on scenarios or different parametrizations. \citet{matthes_impact_2019} explores how \glspl{minram} affect exchange and the number of contingencies, \citet{marjanovic_impact_2018} analyses prices and redispatch quantity for the planned expansion of the flow-based region from \gls{cwe} to the CORE region\footnote{The CORE region extends \gls{cwe} by Austria, the Czech Republic, Hungary, Poland, Romania, Slovakia and Slovenia.}, \citet{wyrwoll_impact_2018} quantifies the effect of security margins on net-positions and generation schedules and \citet{schonheit_impact_2020} the impact of \glspl{gsk} on \glspl{cnec}. 

All papers provide valuable contributions to the field in illustrating the impact of parametrization, but focus on the formal composition of the parameters and do not attribute the dimension of the parametrization that allows to decide on the effect of the parametrization, namely the trade-off between capacity allocation and congestion management. A notable exception is \citet{schonheit_impact_2020} that not only provides transparency by an open modelling approach but explicitly provides insight in the effect of a policy decision influencing an otherwise formal method. The authors also point out differences in how \glspl{minram} can be included in the modelling process. 
\\\\
Extending on exiting research, this paper contributes to academic studies on \gls{fbmc} in different ways: 
\begin{enumerate}
    \item It numerically shows the impact of prioritizing capacity allocation over congestion management in \gls{fbmc}.
    \item The case-study covers full \gls{cwe} for the target year, explicitly covering the medium term influx of intermittent generation and its effect on the efficiency of \gls{fbmc}. 
    \item All data and methods are provided open (under open licence) and accessible (tested and documented) as part of the \glsreset{pomato}\gls{pomato} described in \citet{weinhold_power_2020} and dedicated data processing in \emph{PomatoData} \citep{weinhold_pomatodata_2021}.
\end{enumerate}

\section{Model and Case Study}\label{sec:cwe-model}

The numerical experiments are conducted using the electricity market model \gls{pomato} proposed in \citet{weinhold_power_2020} which was created to model zonal electricity markets and explicitly synthesize the \gls{fbmc} process. The used formulation follows the description in \citet{weinhold_uncertaintyaware_2021} 
and aligns with the process shown in Figure \ref{fig:cwe-fb_process}. Therefore, the model description in this section is limited to a brief description, the full formal description can be found in \ref{app:cwe-formulation}.

Equations~\eqref{eq:cwe-general_dispatch} represent a high level description of the economic dispatch problem that that is used in this study:

\begin{subequations}
\begin{align}
  \min \text{ OBJ} &= \sum \text{COST GEN} + \text{COST CURT} + \text{COST CM} \label{eq:cwe-obj}\\ 
  \text{s.t. }& \nonumber\\
  & \text{Cost Definition} \label{eq:cwe-cost_def} \\
  & \text{Generation Constraints} \label{eq:cwe-generation}\\ 
  & \text{Storage Constraints} \label{eq:cwe-storage}\\
  & \text{Energy Balances} \label{eq:cwe-eb} \\
  & \text{Network Constraints}\label{eq:cwe-network}.
\end{align}
\label{eq:cwe-general_dispatch}%
\end{subequations}%

Each step, the \textit{D-2 capacity forecast (basecase)}, \textit{D-1 day-ahead market clearing} and \textit{D-0 congestion management} utilize the same economic dispatch problem \eqref{eq:cwe-general_dispatch} that finds the most cost effective allocation of generation capacities, defined by generation cost $\text{COST GEN}$ and cost for curtailment of intermittent generation $\text{COST CURT}$, to satisfy demand subject to generation capacity \eqref{eq:cwe-generation} and storage \eqref{eq:cwe-storage} constraints. 
Electricity is balanced for each network node in nodal net-injections and each market area in zonal net-positions \eqref{eq:cwe-eb}, which can be constrained by transport constraints \eqref{eq:cwe-network} that depend on the step in the flow-based process. 

Specifically, the basecase is calculated with linear power flow constraints on nodal net-injections to ensure feasibility on all lines in the network. The day-ahead stage is modeled with transport constraints on net-positions as described in Section \ref{sec:cwe-modeling_fbmc} and Equation \eqref{eq:cwe-F_zonal} for the zones participating in FBMC, other zones are constraint with bilateral \glspl{ntc}. Transmission on \glsreset{hvdc}\gls{hvdc} lines is always included as a decision variable, as \gls{hvdc} lines are considered active network elements. 

Congestion management, similarly to the basecase, solves a nodal market however with additional cost for congestion management $\text{COST CM}$, i.e. redispatch -- deviations from generation schedules of the day-ahead market stage and only conventional, non-storage plants within the flow-based region are considered for redispatch. Curtailment from the market stage persists in congestion management, but can be further increased to maintain feasible power flows as part of the congestion management. 

The resulting total cost reflect cost for generation as well as cost to ensure feasibility in the network. While cost based redispatch is desired, the \textit{D-0 congestion management} should prioritize network feasibility. Therefore the cost parameters for curtailment and redispatch are imposed, that are additive to the generation cost. 
Both are chosen based on average cost for Germany \citep{bundesnetzagentur_monitoring_2019}, with around 25\euro\, per MWh redispatch and 100\euro\, per MWh curtailment. 

\subsection{Input Data}\label{ssec:cwe-input_data}

\begin{figure}[H]
  \centering
    \includegraphics[width=0.95\textwidth]{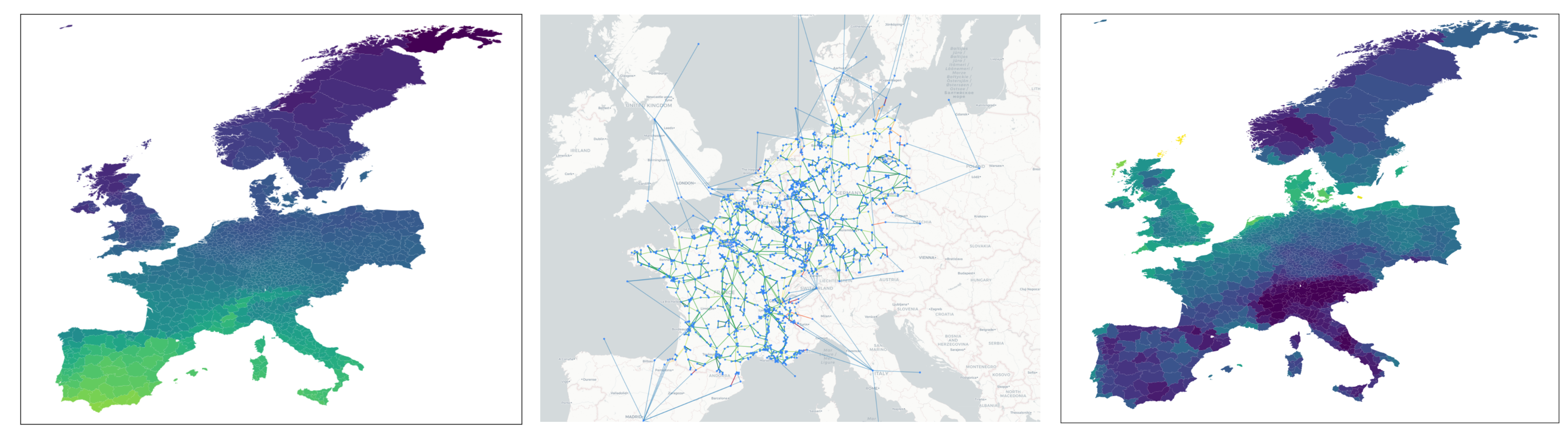}
  \caption[The geographical scope of the model application.]{The geographical scope of the model application and mean solar (left) and wind (right) availability and the transmission network (middle).}
  \label{fig:cwe-model_scope}
\end{figure}

The model application covers the \gls{cwe} region, as depicted in Figure~\ref{fig:cwe-model_scope}, for the years 2020 and 2030. Zones part of the \gls{cwe} are modeled with transmission network and neighboring countries are modeled as single nodes. 

The required data is extensive and collected as well as processed using different contributions by the open-data community and the ENTSO-E Transparency Platform. It uses the \textit{Open Data Portal} of \gls{ffe} \citep{ebner_regionalized_2019} for geo-information and regionalized \gls{res} potentials. 
Geo-information is collected based on the standardised \gls{nuts} data of the European Union, that divides countries in standardised regions and sub-regions and allows for geo-referenced data collection. \gls{nuts}-level 0 defines countries and higher values indicate higher resolution.  
The \textit{atlite} package for availability timeseries on \gls{nuts}-3 level \citep{hofmann_atlite_2021} and hydro storage inflows \citep{liu_highresolution_2019} using the \textit{HydroBASINS} database \citep{lehner_global_2013}.
Dispatchable generation capacities come from the \textit{JRC Hydro-power plants database} \citep{felice_jrc_2021} for hydro generation capacities and the \textit{Open Power System Data Project} \citep{weibezahn_opsd_2018} for conventional and nuclear power plants. 

The underlying network data originates from the updated fork of the \textit{GridKit} ENTSO-E gridmap extract \citep{wiegmans_gridkit_2016} part of well as few specific expansions that were not included in the data, like the already online ``Redwitz-Altenfeld'' and ``Vieselbach-Lauchstädt'' connections. The process of estimating specific line parameters is part of the model documentation\footnote{See: \url{pomato.readthedocs.io/en/latest/line_parameters.html}}.

\begin{figure}[H]
    \centering
    \includegraphics[width=0.95\textwidth]{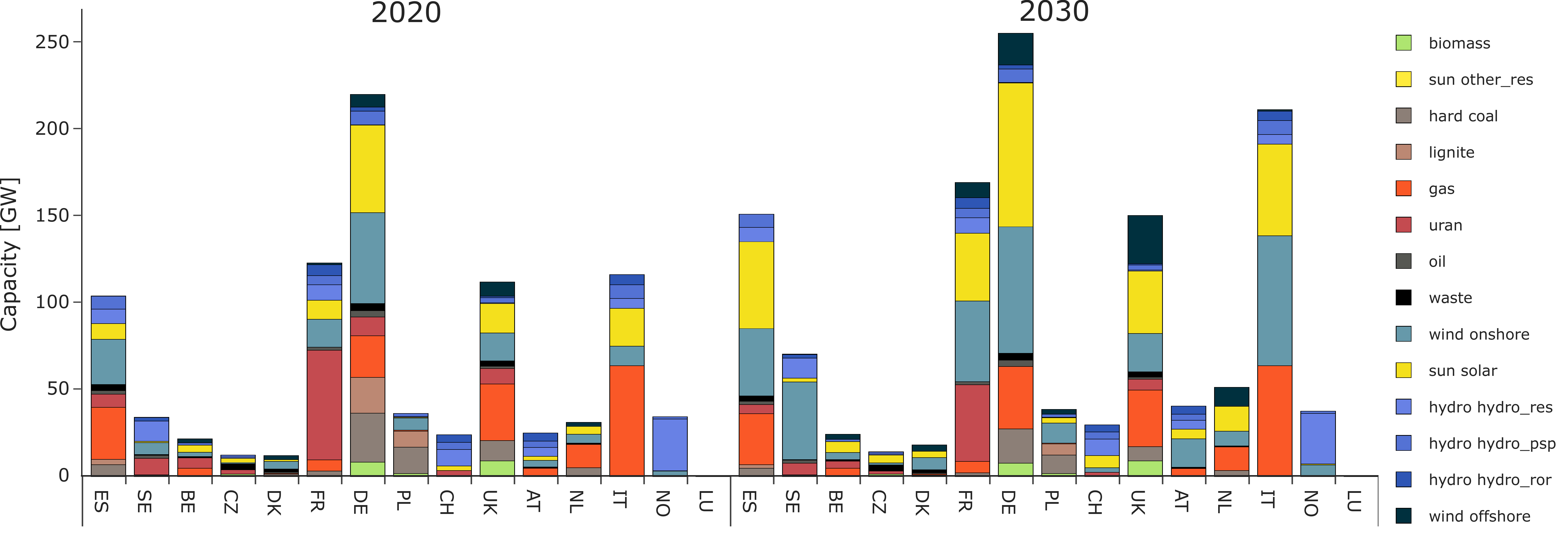}
    \caption{Installed capacities per country and year.}
    \label{fig:cwe-installed_capacities}
\end{figure}

Nodal demand is derived from zonal load published on the \textit{ENTSO-E Transparency Platform}\footnote{See: \url{transparency.entsoe.eu}} using sector specific standard load profiles and \gls{nuts}-3 data on energy consumption, and gross value added \citep{kunz_electricity_2017}. Additional data from \textit{ENTSO-E Transparency Platform} that is used are scheduled commercial exchanges to derive \glspl{ntc} for non-\gls{cwe} bidding zones and weekly storage levels. The first are used to derive \glspl{ntc} for zones neighboring the flow-based region and the latter for rolling horizon model execution that limits the model's foresight. 

Installed capacities of wind and solar in 2030 are obtained from pathway optimization of the European energy system using \textit{AnyMOD} \citep{goke_graphbased_2021}. In contrast to the values part of publications \citet{goke_accounting_2021, hainsch_make_2020} that depict a 2040 fully decarbonized system, the values in this study are obtained for 2030 with the latest ENTSO-E ``sustainable transition'' scenario as a lower bound for renewable capacity and an upper bound for thermal capacities \citep{entsog_tyndp_2020}. The wind and \glsreset{pv}\gls{pv} capacities are distributed based on the \gls{ffe} potentials. Decommissioning of conventional and fissile generation capacities is based on plant lifetime and national energy and climate plans. However, plant specific commissioning and decommissioning of individual plants proves challenging and therefore the used data-set includes modest changes to the conventional and nuclear capacity and the two target years are mostly defined by the regionalized increase in wind and \gls{pv} capacities. Table~\ref{tab:cwe-decomm} shows the decommissioning of lignite, hard coal and nuclear generation capacities between the 2020 and 2030 scenario.

\begin{table}[htbp]
  \centering
  \caption{Decommissioning of lignite, hard coal and nuclear generation capacities in GW per country between 2020 and 2030.}
    \begin{tabular}{lrrrrrrrrr}
    \toprule
          & \multicolumn{1}{l}{BE} & \multicolumn{1}{l}{CH} & \multicolumn{1}{l}{DE} & \multicolumn{1}{l}{ES} & \multicolumn{1}{l}{FR} & \multicolumn{1}{l}{NL} & \multicolumn{1}{l}{PL} & \multicolumn{1}{l}{SE} & \multicolumn{1}{l}{UK} \\
    \midrule
    lignite & 0     & 0     & 20.6  & 0.92  & 0     & 0     & 2.78  & 0     & 0 \\
    hard coal & 0     & 0     & 8.47  & 1.96  & 0.88  & 1.54  & 4.58  & 0.04  & 3.53 \\
    uran  & 1.78  & 1.03  & 10.8  & 2.27  & 18.94 & 0     & 0     & 2.86  & 2.68 \\
    \bottomrule
    \end{tabular}%
  \label{tab:cwe-decomm}%
\end{table}%

The weather year for both 2020 and 2030 was chosen to be 2019. Therefore, 2019 timeseries for load and availability of intermittent generation are used as well as 2019 commercial exchanges used to derive static \glspl{ntc} and weekly storage levels. 

The final data-set is compiled using the complementary PomatoData tool of \gls{pomato} \citep{weinhold_pomatodata_2021} that includes all data processing and documents data origin, thereby provides the required accessibility and compatibility to the data and are published supplementary to the paper.

\subsection{Parametrization of the flow-based parameters}\label{ssec:cwe-fb_parameters}

The aim of this study is to quantify the effect of less restrictive capacity allocation on the resulting congestion management. Therefore the parametrization of the flow-based parameters remains very close to the official documentation \citep{50hertz_documentation_2020} for \gls{gsk}, \gls{cnec} selection and the implementation of the \gls{minram} criterion.

For all scenarios a \textit{Pro-Rata} \gls{gsk} is used. This \gls{gsk} weights nodal participation factors based on the scheduled power output of dispatchable generation capacities and is the basis for most \glspl{gsk} currently in use \citep{50hertz_documentation_2020} and has proven to be effective \citep{schonheit_impact_2020}. For the construction of the zonal \gls{ptdf}, \glspl{cne} are selected based on a 5\% threshold in the zone-to-zone \gls{ptdf}, with the exception of one scenario where only cross-border lines are considered critical. Contingencies are selected based on a 20\% sensitivity threshold of a contingency towards the \gls{cne}, i.e. contingencies are considered if in case of an outage more than 20\% of the load is diverted to the respective \gls{cne}. 

\begin{figure}[H]
    \centering
    \includegraphics[width=0.95\textwidth]{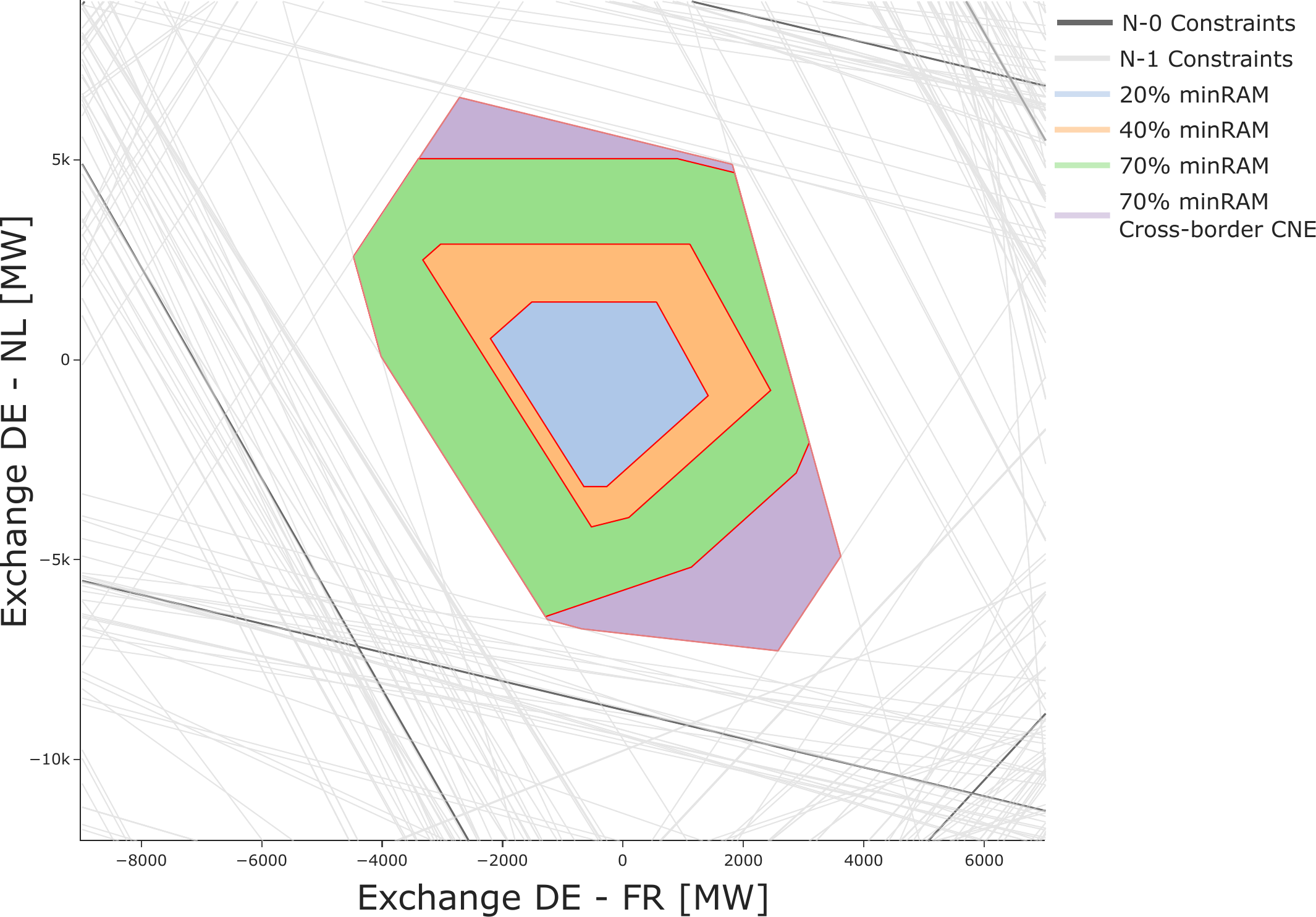}
    \caption[Visualization of the day-ahead CA for exchange between Germany-France.]{Visualization of the day-ahead CA for exchange between Germany-France (x-axis) and Germany-Netherlands (y-axis) depending on the chosen \gls{minram} configuration.}
    \label{fig:cwe-fb_domain}
\end{figure}

The \gls{minram} criterion is enforced by setting the RAM to $\max(RAM, \overline{f} \cdot \text{minRAM})$ in Equation \eqref{eq:cwe-F_zonal} and the model is run for different \gls{minram} values of 20\%, 40\% and 70\%. In addition the 70\% is also run with only cross-border lines as \gls{cne} indicated as \textit{70\% (only CB)} in the following result tables. The effect on the day-ahead capacity allocation is visualized in a flow-based domain for the exchange between Germany-France and Germany-Netherlands in Figure \ref{fig:cwe-fb_domain}. 

The four configurations differ in the permissiveness of commercial exchange and we can expect higher trading volumes with higher \gls{minram} requirements. Additionally, in order to further limit the impact of internal congestions on cross-border exchange \citep{europeancommission_commission_2019} the fourth scenario only considers cross-border lines as \glspl{cne}, thus further relaxing the commercial exchange domains. 

The final 2030 data-set is composed of 5458 generators, that includes 2236 wind/\gls{pv} plants and 953 storages. The network is made up of 1663 nodes, 3276 lines and 100 \gls{hvdc} lines. The model size is substantial and to alleviate the computational effort, the model is solved for every 7th week of the year, 8 weeks in total, with the electricity market model \gls{pomato} \citep{weinhold_power_2020}. Therefore the model remains tractable on standard pc hardware, all results are obtained using a AMD Ryzen 7 CPU, 32GB of memory and the Gurobi solver \citep{gurobioptimizationllc_gurobi_2018}.   

\section{Model Results}\label{sec:cwe-results}

\begin{figure}[H]
  \centering
    \includegraphics[width=0.95\textwidth]{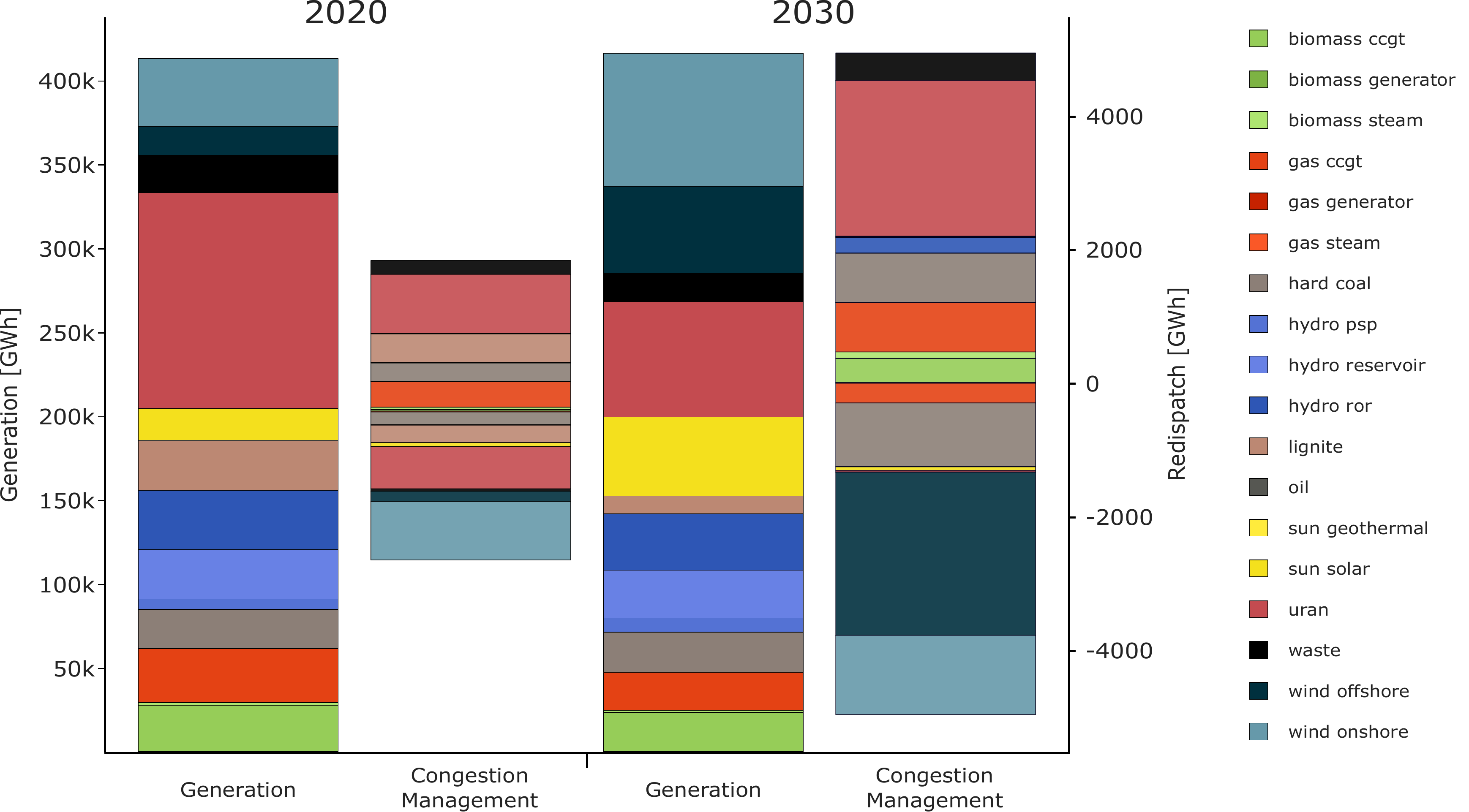}
  \caption[Total generation and congestion management by fuel-type for the \textit{70\% minRAM}.]{Total generation and congestion management by fuel-type for the \textit{70\% minRAM} configuration in 2020 (left) and 2030 (right).}
  \label{fig:cwe-generation_overview}
\end{figure}

As described in Section~\ref{sec:cwe-model} \gls{fbmc} is modeled in three dedicated steps. From a basecase the day-ahead market result is obtained using flow-based parameters following the four configurations regarding the permissiveness of the day-ahead trading domain and subsequent congestion management. The model results are evaluated after congestion management and therefore include generation cost from the market- and congestion management stage, as well as additional cost for redispatch and curtailment. Figure~\ref{fig:cwe-generation_overview} presents the resulting dispatch for the \textit{70\% minRAM} configuration in 2020 and 2030 and illustrates the overall system states with a renewable share of 43\% in 2020 and 66\% in 2030 as well as the volumes and composition of congestion management.

As the primary result metric, the system cost indicate the efficiency of the market design and when decomposed into cost for generation, curtailment and redispatch show the relation between capacity allocation and congestion management. 
Table~\ref{tab:cwe-cost} shows lower system cost in 2030 due to an increased share of intermittent renewable generation and decreasing generation cost with more permissive commercial exchange domains. 
Cost for congestion management are the highest for the \textit{20\% minRAM} configuration, indicating overly conservative trading domain and in combination with the least effective use of generation capacities, results in highest total cost. 
With higher commercial exchange capacities curtailment increases and redispatch quantities decrease. The lowest total cost align with the lowest congestion management volume, depicted in Table~\ref{tab:cwe-cm_quantity}, in the \textit{70\% minRAM} configuration for 2020 and the less constraint \textit{70\% minRAM (only CB)} configuration in 2030. 

\begin{table}[h]
\centering
\caption{Total cost for generation, curtailment and redispatch in mio.~\euro}\label{tab:cwe-cost}
\begin{tabular}{llllll} \toprule
                      &                                  & \multicolumn{4}{c}{minRAM configuration}      \\ \cmidrule(l){3-6} 
                      &             & 20\%        & 40\%        & 70\%          & 70\% (only CB)        \\ \midrule
\multirow{4}{*}{2020} & Generation  & 9,837       & 9,758       & 9,718         & 9,704                 \\
                      & Curtailment & 104         & 108         & 111           & 117                   \\
                      & Redispatch  & 129         & 94          & 84            & 98                    \\ \cmidrule(l){2-6} 
                      & Total       & 10,070      & 9,960       & 9,913         & 9,919                 \\ \midrule
\multirow{4}{*}{2030} & Generation  & 6,926       & 6,905       & 6,888         & 6,864                 \\
                      & Curtailment & 559         & 560         & 567           & 588                   \\
                      & Redispatch  & 180         & 160         & 155           & 146                   \\ \cmidrule(l){2-6} 
                      & Total       & 7,665       & 7,625       & 7,610         & 7,598                 \\ \bottomrule
\end{tabular}
\end{table}

Overall, the best configuration in 2020 represents a 1.56\% cost reduction and 27.84\% reduction in congestion management in comparison to the \textit{20\% minRAM} configuration. In 2030 the total cost reduction is 0.87\% and congestion management is reduced by 8.34\%.

\begin{table}[h]
\centering
\caption{Congestion management volumes in TWh}\label{tab:cwe-cm_quantity}
\begin{tabular}{@{}llllll@{}} \toprule
                      &             & \multicolumn{4}{c}{minRAM configuration} \\ \cmidrule(l){3-6} 
                      &             & 20\%   & 40\%   & 70\%  & 70\% (only CB) \\ \midrule
\multirow{3}{*}{2020} & Curtailment & 1.05   & 1.09   & 1.11  & 1.17           \\
                      & Redispatch  & 5.2    & 3.77   & 3.4   & 3.94           \\ \cmidrule(l){2-6} 
                      & Total       & 6.25   & 4.86   & 4.51  & 5.11           \\ \midrule
\multirow{3}{*}{2030} & Curtailment & 5.6    & 5.6    & 5.68  & 5.89           \\
                      & Redispatch  & 7.23   & 6.43   & 6.22  & 5.87           \\ \cmidrule(l){2-6} 
                      & Total       & 12.83  & 12.03  & 11.9  & 11.76          \\ \bottomrule
\end{tabular}
\end{table}

\begin{figure}[H]
  \centering
    \includegraphics[width=0.95\textwidth]{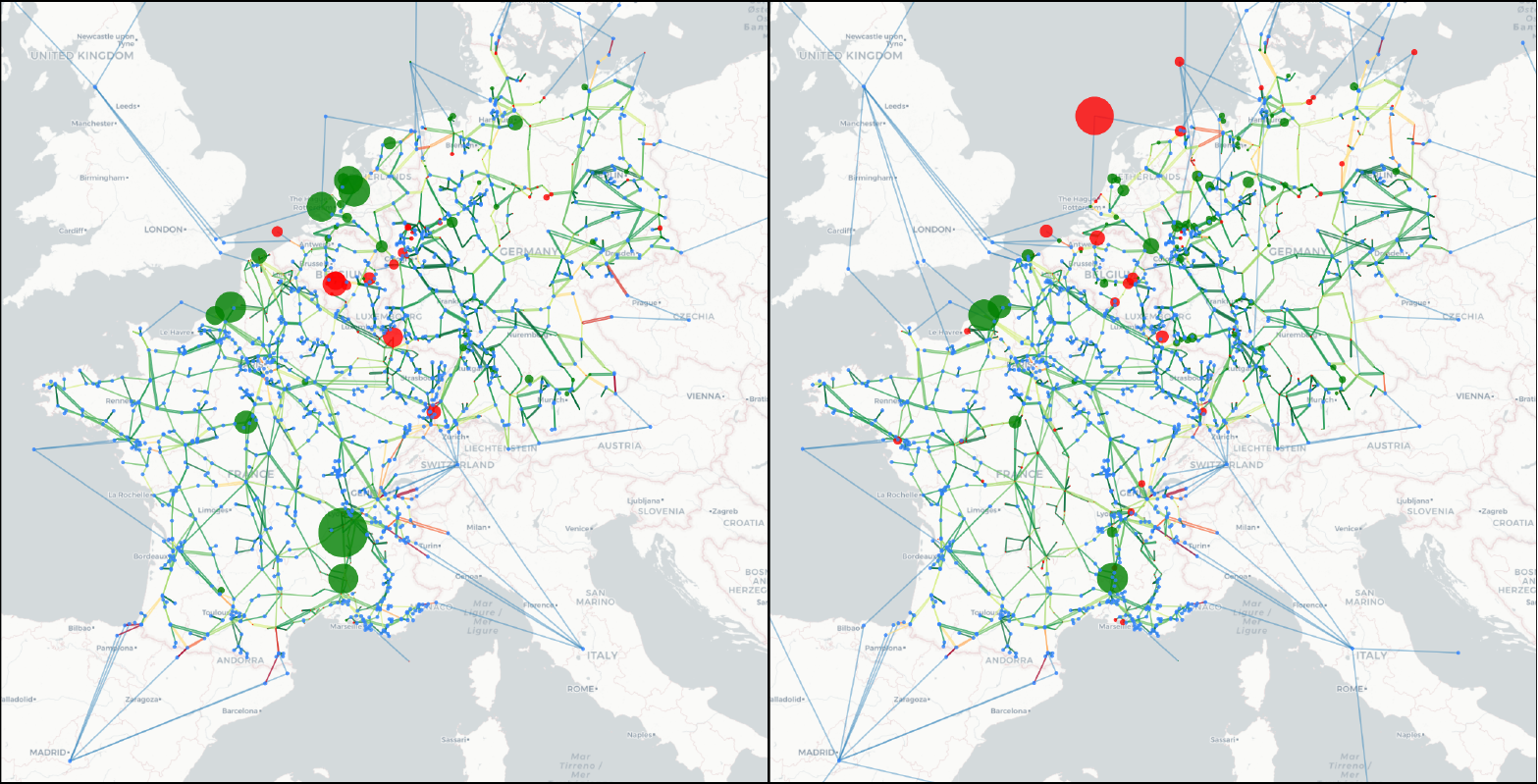}
  \caption[Difference in congestion management for 2020 (left) and 2030 (right).]{Difference in congestion management for 2020 (left) and 2030 (right) between the \textit{20\% minRAM} and \textit{70\% minRAM} configurations. Increased congestion management with \textit{70\% minRAM} is indicated in red and less congestion management in green.}
  \label{fig:cwe-geoplot_redispatch}
\end{figure}

In order to evaluate the impact on congestion management Figure~\ref{fig:cwe-geoplot_redispatch} shows the difference in congestion management between the \textit{20\% minRAM} and \textit{70\% minRAM} configurations for 2020 and 2030. For both years less restricted commercial exchange causes more congestion management along the \gls{cwe} border region, which is out-weight by less congestion management generally more within the bidding zones. The impact of the substantial increases in offshore wind capacities in the north sea is also clearly visible.  

A core metric of the European \gls{iem} is price convergence, as it quantifies increased market efficiency and competition that should result in lower prices \citep{europeancommission_directive_2019}.
Table \ref{tbl:price} shows different compositions of the model endogenous electricity prices $\lambda$ resulting from the dual variable of the energy balances. 
For the day-ahead (D-1) stage $\lambda$ is the average zonal price for the \gls{cwe} bidding zones and $\Delta \lambda$ the average difference in zonal price between the \gls{cwe} bidding zones. For congestion management (D-0), with nodal restrictions present, nodal marginal prices in congestion management $\lambda$ are averaged. The marginal in congestion management indicates the additional nodal cost after congestion management, meaning after network feasibility is achieved. 


\begin{table}[htbp]
  \centering
  \caption[Average price $\lambda$ and price differences $\Delta \lambda$.]{Average price $\lambda$ and price differences $\Delta \lambda$ in- and between \gls{cwe} bidding zones in terms of year and \gls{minram} configuration.}\label{tbl:price}
    \begin{tabular}{clllllllll}
    \toprule
          &       & \multicolumn{2}{c}{20\%} & \multicolumn{2}{c}{40\%} & \multicolumn{2}{c}{70\%} & \multicolumn{2}{c}{70\% (only CB)} \\
          &       & \multicolumn{1}{l}{$\lambda$} & \multicolumn{1}{l}{$\Delta \lambda$} & \multicolumn{1}{l}{$\lambda$} & \multicolumn{1}{l}{$\Delta \lambda$} &  \multicolumn{1}{l}{$\lambda$} & \multicolumn{1}{l}{$\Delta \lambda$} & \multicolumn{1}{l}{$\lambda$} & \multicolumn{1}{l}{$\Delta \lambda$} \\
    \midrule
    \multirow{2}[1]{*}{2020} & D-1 & 50.02 & 24.41 & 44.68 & 22.4  & 41.61 & 21.23 & 39.08 & 17.06 \\
          & D-0 & 47.18 &       & 48.13 &       & 48.46 &       & 47.88 &  \\ \midrule
    \multirow{2}[1]{*}{2030} & D-1 & 43.47 & 31.73 & 42.02 & 24.62 & 41.3  & 22.49 & 37.58 & 19.46 \\
          & D-0 & 54.39 &       & 55.48 &       & 56.32 &       & 57.67 &  \\
    \bottomrule
    \end{tabular}%
  \label{tab:cwe-addlabel}%
\end{table}%

Similar to the system cost results in Table~\ref{tab:cwe-cost}, average prices decrease between 2020 and 2030 due to the influx of low cost generation.
For both 2020 and 2030 the average prices decrease with less restricted commercial exchange domains. Additionally, price differences decrease with more exchange capacities with 2030 showing, on average, price differences at lower average prices.

\begin{figure}[H]
    \centering
    \includegraphics[width=1\textwidth]{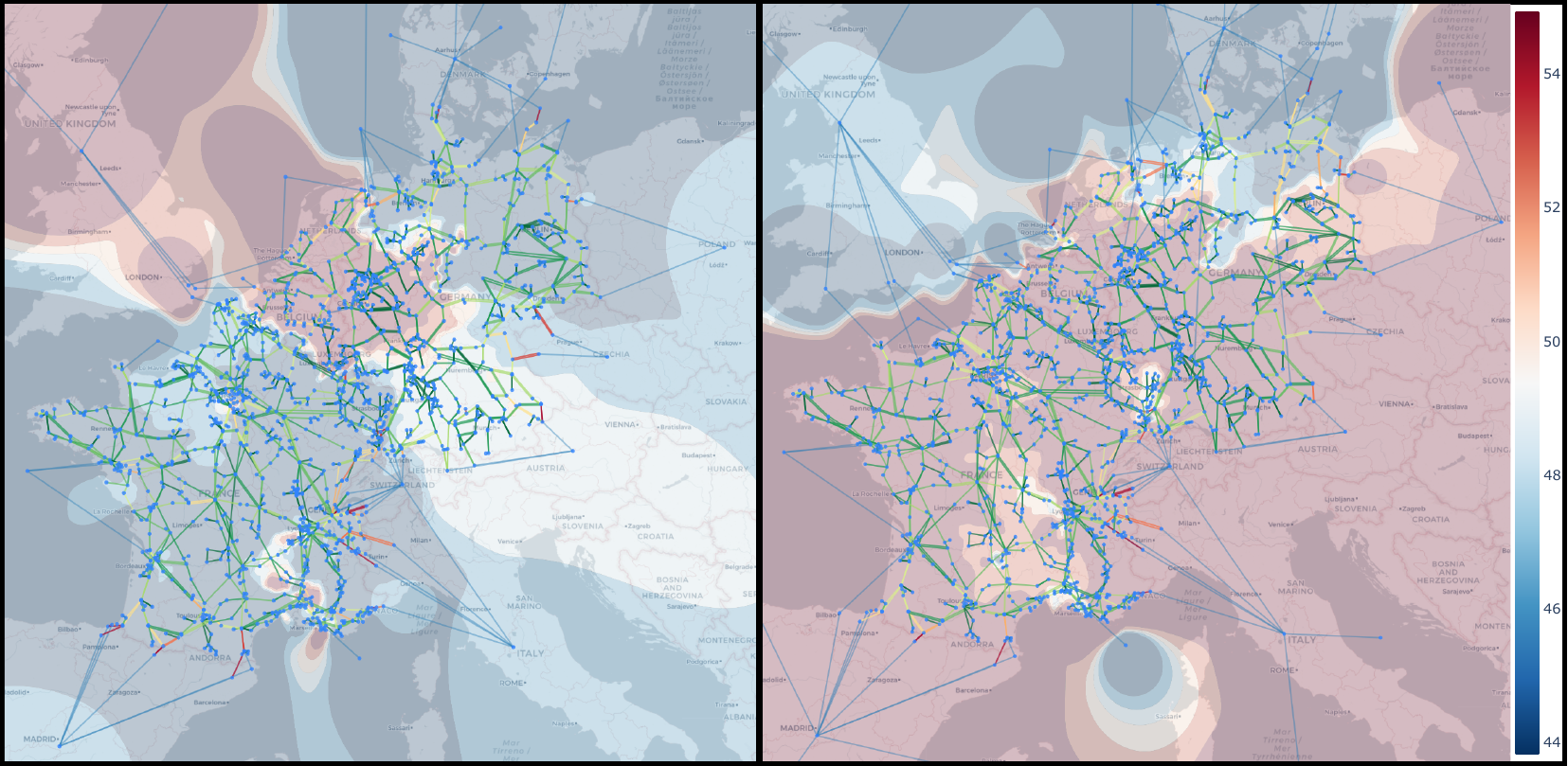}
    \caption{Average nodal marginal cost for congestion management in 2020 (left) and 2030 (right) for the \textit{70\% minRAM} configuration.}
    \label{fig:cwe-price_map}
\end{figure}

Results of the average marginal cost for electricity after congestion management show generally higher values in 2030, reflecting the increased congestion management volumes. This is visualized in Figure~\ref{fig:cwe-price_map}, that shows the average nodal prices after congestion management by the dual on the energy balances as a contour plot for the modeled region for the \textit{70\% minRAM} configuration. 

The plot verifies the overall increased congestion management effort in 2030. Also, given the distribution of \gls{res} capacities by potential, 2030 shows fairly even contour in the \gls{cwe} region. 

Interestingly, between the \gls{fbmc} configurations, the values negatively correlate to the congestion management volumes, with higher prices for lower overall volumes. Thereby, higher average price after congestion management indicates that additional nodal load is more often subject to network constraints, or indicate higher average network utility. 

\section{Conclusions}

The presented results show the effect of \gls{fbmc} configurations that, in line with the regulators perspective, provide higher commercial exchange capacities to the market. The modeled configurations are chosen to represent the status-quo and the proposed target by 2030, modeled through different \gls{minram} requirements and choice of critical network elements.

More permissive configurations provide lower generation cost, as the markets are less constrained. 
For 2020, the trade-off between capacity allocation and congestion management is visible. Here both, too constrained as well as too permissive flow-based parameter configurations lead to higher overall cost. The least cost solution is provided by the \textit{70\% minRAM} configuration.  
For 2030 the most permissive configuration \textit{70\% minRAM (only CB)} provided the least over cost and congestion management. 

In terms of price convergence, the results show that indeed more permissive exchange capacities lead to higher levels of price convergence in both 2020 and 2030. However, 2030 shows an overall lower price convergence than 2020 at lower overall price level, indicating higher price volatility in 2030. 
Analysis of marginal cost in congestion management also reflect the overall higher volumes of congestion management. Additionally, the comparison between the \gls{fbmc} configurations leads to the conclusion that more permissive flow-based domains lead to a higher network utilization. 

This paper provides comprehensive insights into the effect of policy relevant considerations as part of \gls{fbmc}. This includes the status-quo as well as the mid-term future. Indeed, enforcing larger commercial exchange capacities lead to increased price convergence, however potentially with increased congestion management. The results also indicate that larger commercial exchange domains are favourable in high shares of intermittent renewable generation. However, the 2030 scenario also shows high levels of congestion management, that indicate the systems inability to adequately make use or transport excess intermittent renewable generation. The extensive use of curtailment as part of congestion management, despite the high cost, also indicates that the homogeneous distribution makes curtailment very effective tool for congestion management. 
The 2030 scenario in this study differs from the status-quo mainly in terms of the renewable capacities and the \gls{hvdc} projects of the immediate future. Thereby the effect of higher shares of intermittent renewable generation becomes visible. 

The methods and data that are used in this research are open and accessible and aim to provide transparency and sustainability to the research and results of this study as well prove useful for future work. The proposed data-set that accompanies the used electricity market model \gls{pomato} \citep{weinhold_power_2020} allows for detailed and in-depth analysis of large-scale electricity system in respect to market design and the involved policy decisions. 

For future studies, the impact of means to reduce excess generation, flatten load and increase system flexibility via sector coupling, prosumage or demand response would provide relevant avenues for extensions. The expected decrease in congestion management as well as additional flexibility in terms of network operation would represent valuable insights. 

\section*{Aknowledements}
The author gratefully acknowledges the support by the German Federal Ministry for Economic Affairs and Energy (BMWi) in the project “Modellierung (De-)Zentraler Energiewenden: Wechselwirkungen, Koordination und Lösungsansätze aus systemorientierter Perspektive” (MODEZEEN, 03EI1019B)








\newpage

\appendix 

\section{Nomenclature}

\begin{longtable}{p{.10\textwidth} p{.80\textwidth}}
	\caption{Nomenclature.}\label{tab:cwe-nomenclature} \\ \toprule
	\endfirsthead
	\caption*{Nomenclature (continued).} \\ \toprule
	\endhead
	\multicolumn{2}{l}{\textbf{A. Sets}} \\ \midrule
	$\set{T}$ 	& Set of time steps within the model horizon.  \\
	$\set{G}$ 	& Set of generators.  \\
	$\set{R}$ 	& Subset $\set{R} \subset \set{P}$ of intermittent generators.  \\
	$\set{N}$ 	& Set of network nodes. \\
	CNEC        & Set of critical network elements and contingencies. \\
	$\set{Z}$ 	& Set of bidding zones. \\
	\midrule
	\multicolumn{2}{l}{\textbf{B. Variables}} \\ \midrule
	$G_t$ 	& Active power generation at $t\in\set{T}$ indexed by $G_{g,t}, g\in\set{G}$ \\
	$C_t$ 	& Curtailment $t\in\set{T}$. \\
	$I_t$ 	& Active nodal power injection at $t\in\set{T}$. \\
	$NP_t$ 	& Zonal net-positions at $t\in\set{T}$. \\
	$F^{dc}_t$ 	& Flow on HVDC lines at $t\in\set{T}$. \\
	$EX_t$ 	& Bilateral exchange at $t\in\set{T}$ indexed by $EX_{t,z,z'}, z,z'\in\set{Z}$ \\
	$L_t$ 	& Storage level at $t\in\set{T}$ indexed by $L_{t,s}, s\in\set{ES}$ \\
	$D_t$ 	& Storage demand/charging at $t\in\set{T}$ indexed by $D_{t,s}, s\in\set{ES}$ \\
	$G^{red}_t$ & Active power redispatch at $t\in\set{T}$. \\
	\midrule
	\multicolumn{2}{l}{\textbf{C. Parameters}} \\ \midrule
	$d_t$ 	& Nodal load at $t\in\set{T}$.	\\
	$r_t$ 	& Available intermittent generation at $t\in\set{T}$	\\
	$\mathrm{m}$ 	& Mapping of generators/load (subscript) to nodes/zones (superscript).	\\
	$\overline{g}$ 	& Upper generation limit.	\\
	$\overline{d}$ 	& Upper storage charging limit indexed by $\overline{d}_{t,s},s\in\set{ES}$	\\
	$\eta$ 	& Storage charging efficiency indexed by $\eta_s,s\in\set{ES}$	\\
	$\overline{l}$ 	& Upper storage capacity indexed by $\overline{l}_{t,es},es\in\set{ES}$	\\
	$\overline{f}$ 	& Maximum line flow.	\\
	$\overline{f^{dc}}$ 	& Maximum dc-line flow.	\\
	$ntc$ 	        & Net-transfer capacity indexed by $ntc_{z,zz}, z,z'\in\set{Z}$	\\
	$g^{da}$ 	& Day-ahead scheduled generation at $t\in\set{T}$. \\
	$c^{da}$ 	& Day-ahead scheduled curtailment at $t\in\set{T}$.\\
	\bottomrule
\end{longtable}

\section{Model Formulation}\label{app:cwe-formulation}

Equations~\ref{eqs:cwe-general_economic_dispspatch} formulate the economic dispatch problem that finds the least cost dispatch to satisfy demand for each timestep $t\in\set{T}$. 

\begin{subequations}\label{eqs:cwe-general_economic_dispspatch}%
    \begin{align}
        \min \quad & \sum_{t\in\mathcal{T}}c(G_t) + c(C_t) && \label{eq:cwe-app-obj}\\
        \text{s.t. } \quad &0 \leq G_t \leq \overline{g} && \forall t \in \set{T} \label{eq:cwe-app-capacity_g}\\
        &0 \leq C_t \leq r_t && \forall t \in \set{T} \label{eq:cwe-app-capacity_c}\\
        &L_{t, s} = L_{t-1, s} - G_{t,s} + \eta_s D_{t,s} &&\forall s \in \set{ES}, t \in \set{T}\label{eq:cwe-app-storage_balance}\\
        &0 \leq D_t \leq \overline{d} && \forall t \in \set{T} \label{eq:cwe-app-capacity_d}\\
        &0 \leq L_t \leq \overline{l} && \forall t \in \set{T}, \label{eq:cwe-app-capacity_l} \\
        & -\overline{f^{dc}} \leq F^{dc}_t \leq \overline{f^{dc}} && \forall t \in \set{T}, \label{eq:cwe-app-capacity_dc} \\
        &\mathrm{m}_g^n G_t + \mathrm{m}_r^n ( r_t - C_t) && \nonumber \\
        &\qquad - \mathrm{m}_g^n D_t + A^{dc} \cdot F_t^{dc} - d_t = I_t && \forall t \in \set{T} \label{eq:cwe-app-eb_nodal}\\  
        &\mathrm{m}_g^z G_t + \mathrm{m}_r^z ( r_t - C_t)  && \nonumber \\
        &\qquad - \mathrm{m}_g^z D_t - m_d^z d_t = NP_t && \forall t \in \set{T} \label{eq:cwe-app-eb_zonal} \\
        & NP_{t,z} = \sum_{z'\in\mathcal{Z}} \EX_{t,z,z'} - \EX_{t,z',z} && \forall t\in\mathcal{T}, \forall z\in \mathcal{Z} \label{eq:cwe-app-ex}  \\
        &e^T I_t = 0 &&\forall t \in \set{T} \label{eq:cwe-app-balance}
    \end{align}
\end{subequations}

The objective function~\eqref{eq:cwe-app-obj} minimizes generation cost, given by a cost function $c(\cdot)$ and the vector of hourly generation levels $G_t$ and curtailment $C_t$. 
Equations \eqref{eq:cwe-app-capacity_g} and \eqref{eq:cwe-app-capacity_c} provide bounds to generation and curtailment based on the installed capacity $\overline{g}$ and available capacity $r_t$.

Storages are modeled using a storage balance \eqref{eq:cwe-app-storage_balance} and bounds on storage charging \eqref{eq:cwe-app-capacity_d} and storage level \eqref{eq:cwe-app-capacity_l}. Note, the storage balance requires parametrization of start- and end levels for feasibility. In the application in this study, weekly historic storage levels are used for this purpose.

Generation and load are balanced in nodal injections for each node in \eqref{eq:cwe-app-eb_nodal}
and zonal net-positions for each zones in \eqref{eq:cwe-app-eb_zonal}. The net-position can be expressed as the sum of bilateral exchanges for each zone as per~\eqref{eq:cwe-app-ex}. The positive variable $\EX$ captures bilateral exchange but requires physical connection between zones, otherwise it is fixed to zero.

\gls{hvdc} lines, as active network elements are always part of the economic dispatch problem. Flow on \gls{hvdc} lines $F^{dc}$ is bounds by line's capacity $\overline{f^{dc}}$ in \eqref{eq:cwe-app-capacity_dc} and included in the nodal energy balance with an incidence matrix $A^{dc}$ that maps flows to start- and end nodes.  

The entire system is balanced with constraint \eqref{eq:cwe-app-balance}. 

\subsection{Network Constraints}

As discussed in Section~\ref{sec:cwe-model}, problem \eqref{eqs:cwe-general_economic_dispspatch} can be subject to transport constraints to model the different \gls{fbmc} steps. 

\begin{align}
    I_t \in \set{F}^n &\coloneqq \{ x : \PTDF^n x \leq \overline{f}\} &&\forall t \in \set{T}. \label{eq:cwe-app-F_nodal} \\
    NP_t \in \set{F}^z &\coloneqq \{ x : \PTDF^z_t x \leq RAM_t\} &&\forall t \in \set{T},\label{eq:cwe-app-F_zonal} \\
    EX_t \in \set{F}^{ntc} &\coloneqq \{ x : 0 \leq x \leq ntc \} &&\forall t \in \set{T}. \label{eq:cwe-app-F_ntc}
\end{align}

Equation~\eqref{eq:cwe-app-F_nodal} defines the feasible region for nodal net-injections using the nodal \gls{ptdf} matrix and line capacities $\overline{f}$. 
Equation~\eqref{eq:cwe-app-F_zonal} constrains the zonal net-positions with the flow-based parameters, as described in Section~\ref{sec:cwe-modeling_fbmc}. 
Equation~\eqref{eq:cwe-app-F_ntc} defines bounds on bilateral commercial exchange. 

\subsection{Congestion management}

Congestion management finds the least cost deviation from day-ahead generation schedule $g_t^{da}$ and $c_t^{da}$ that are the decisions on $G_t$ and $C_t$ from the previous market clearing stage while creating network feasibility as per \eqref{eq:cwe-app-F_nodal}.

\begin{subequations}\label{eqs:cwe-redispatch}
    \begin{align}
        C(G^{red}) &= c^{red} \sum_{t \in \set{T}} |G^{red}_t| \label{eq:cwe-redisp_cost}\\
        G_t - g^{da}_t &= G^{red}_t &&\forall t \in \set{T} \label{eq:cwe-redisp_def}\\
        C_t &\geq c^{da} &&\forall t \in \set{T}, \label{eq:cwe-redisp_curtlimit}
    \end{align}
\end{subequations}

Equation~\eqref{eq:cwe-redisp_cost} captures the additional cost that occur for changing the day-ahead generation schedule. Constraints \eqref{eq:cwe-redisp_def} and \eqref{eq:cwe-redisp_curtlimit} bound the allowed deviation from day-ahead schedule for generation and curtailment. 

The dispatch problem for congestion management is therefore:

\begin{align}
    &\min \quad \eqref{eq:cwe-app-obj} + \eqref{eq:cwe-redisp_cost} &&\\
    \text{s.t. } & \eqref{eq:cwe-app-capacity_g} - \eqref{eq:cwe-app-balance} &&\\
    & \eqref{eq:cwe-redisp_def} + \eqref{eq:cwe-redisp_curtlimit} &&
\end{align}

\clearpage
\bibliographystyle{elsarticle-harv}
\bibliography{references_bibtex.bib}

\begin{thebibliography}{39}
\expandafter\ifx\csname natexlab\endcsname\relax\def\natexlab#1{#1}\fi
\providecommand{\url}[1]{\texttt{#1}}
\providecommand{\href}[2]{#2}
\providecommand{\path}[1]{#1}
\providecommand{\DOIprefix}{doi:}
\providecommand{\ArXivprefix}{arXiv:}
\providecommand{\URLprefix}{URL: }
\providecommand{\Pubmedprefix}{pmid:}
\providecommand{\doi}[1]{\href{http://dx.doi.org/#1}{\path{#1}}}
\providecommand{\Pubmed}[1]{\href{pmid:#1}{\path{#1}}}
\providecommand{\bibinfo}[2]{#2}
\ifx\xfnm\relax \def\xfnm[#1]{\unskip,\space#1}\fi
\bibitem[{{50Hertz} et~al.(2020){50Hertz}, {Amprion}, {APG}, {Creos}, {Elia},
  {Rte}, {TenneT} and {TransnetBW}}]{50hertz_documentation_2020}
\bibinfo{author}{{50Hertz}}, \bibinfo{author}{{Amprion}},
  \bibinfo{author}{{APG}}, \bibinfo{author}{{Creos}}, \bibinfo{author}{{Elia}},
  \bibinfo{author}{{Rte}}, \bibinfo{author}{{TenneT}},
  \bibinfo{author}{{TransnetBW}}, \bibinfo{year}{2020}.
\newblock \bibinfo{title}{Documentation of the {{CWE FB MC}} Solution -
  {{Version}} 5.0}.
\newblock \bibinfo{type}{Technical {{Report}}}.
\bibitem[{{ACER}(2016)}]{acer_decision_2016}
\bibinfo{author}{{ACER}}, \bibinfo{year}{2016}.
\newblock \bibinfo{title}{Decision of the {{Agency}} for the {{Cooperation}} of
  {{Energy Regulators No}} 06/2016 of 17 {{November}} 2016: {{On}} the
  {{Electricity Transmission System Operators}}' {{Proposal}} for the
  {{Determination}} of {{Capacity Calculation Regions}}}.
\newblock \bibinfo{type}{Technical Report}.
\bibitem[{{ACER} and {CEER}(2020)}]{acer_annual_2020}
\bibinfo{author}{{ACER}}, \bibinfo{author}{{CEER}}, \bibinfo{year}{2020}.
\newblock \bibinfo{title}{Annual {{Report}} on the {{Results}} of
  {{Monitoring}} the {{Internal Electricity}} and {{Natural Gas Markets}} in
  2019}.
\newblock \bibinfo{type}{Report}.
\bibitem[{{Amprion}(2019)}]{amprion_amprion_2019}
\bibinfo{author}{{Amprion}}, \bibinfo{year}{2019}.
\newblock \bibinfo{title}{Amprion {{Market Report}} 2019 - {{Flow Based Market
  Coupling}}: {{Development}} of the {{Market}} and {{Grid Situation}}
  2015-2018}.
\newblock \bibinfo{type}{Report}.
\bibitem[{{Amprion} et~al.(2011){Amprion}, {APX-ENDEX}, {Belpex}, {Creos},
  {Elia}, {EnBW}, {EPEX SPOT}, {RTE} and {TenneT}}]{amprion_cwe_2011}
\bibinfo{author}{{Amprion}}, \bibinfo{author}{{APX-ENDEX}},
  \bibinfo{author}{{Belpex}}, \bibinfo{author}{{Creos}},
  \bibinfo{author}{{Elia}}, \bibinfo{author}{{EnBW}}, \bibinfo{author}{{EPEX
  SPOT}}, \bibinfo{author}{{RTE}}, \bibinfo{author}{{TenneT}},
  \bibinfo{year}{2011}.
\newblock \bibinfo{title}{{{CWE Enhanced Flow}}-{{Based MC}} feasibility
  report}.
\bibitem[{Boury(2015)}]{boury_methods_2015}
\bibinfo{author}{Boury, J.}, \bibinfo{year}{2015}.
\newblock \bibinfo{title}{Methods for the Determination of Flow-Based Capacity
  Parameters: Description, Evaluation and Improvements}.
\newblock \bibinfo{type}{Master's {{Thesis}}}. KU Leuven.
\bibitem[{{Bundesnetzagentur} and
  {Bundeskartellamt}(2019)}]{bundesnetzagentur_monitoring_2019}
\bibinfo{author}{{Bundesnetzagentur}}, \bibinfo{author}{{Bundeskartellamt}},
  \bibinfo{year}{2019}.
\newblock \bibinfo{title}{Monitoring {{Report}} 2019}.
\newblock \bibinfo{type}{Report}.
\bibitem[{Byers and Hug(2020)}]{byers_modeling_2020}
\bibinfo{author}{Byers, C.}, \bibinfo{author}{Hug, G.}, \bibinfo{year}{2020}.
\newblock \bibinfo{title}{Modeling flow-based market coupling: {{Base}} case,
  redispatch, and unit commitment matter}, in: \bibinfo{booktitle}{2020 17th
  {{International Conference}} on the {{European Energy Market}}}, pp.
  \bibinfo{pages}{1--6}.
\bibitem[{{Directorate General for
  Energy}(2019)}]{directorategeneralforenergy_clean_2019}
\bibinfo{author}{{Directorate General for Energy}}, \bibinfo{year}{2019}.
\newblock \bibinfo{title}{Clean Energy for All {{Europeans}}}.
\newblock \bibinfo{type}{White {{Paper}}}. {European Commision}.
\bibitem[{Ebner et~al.(2019)Ebner, Fiedler, Jetter and
  Schmid}]{ebner_regionalized_2019}
\bibinfo{author}{Ebner, M.}, \bibinfo{author}{Fiedler, C.},
  \bibinfo{author}{Jetter, F.}, \bibinfo{author}{Schmid, T.},
  \bibinfo{year}{2019}.
\newblock \bibinfo{title}{Regionalized {{Potential Assessment}} of {{Variable
  Renewable Energy Sources}} in {{Europe}}}, in: \bibinfo{booktitle}{2019 16th
  {{International Conference}} on the {{European Energy Market}}}, pp.
  \bibinfo{pages}{1--5}.
\bibitem[{{ENTSOG} and {ENTSO-E}(2020)}]{entsog_tyndp_2020}
\bibinfo{author}{{ENTSOG}}, \bibinfo{author}{{ENTSO-E}}, \bibinfo{year}{2020}.
\newblock \bibinfo{title}{{{TYNDP}} 2020: {{Scenario Report}}}.
\newblock \bibinfo{type}{Report}.
\bibitem[{{ETSO}(2001)}]{etso_procedures_2001}
\bibinfo{author}{{ETSO}}, \bibinfo{year}{2001}.
\newblock \bibinfo{title}{Procedures for {{Cross}}-{{Border Transmission
  Capacity Assessments}}}.
\newblock \bibinfo{type}{Technical Report}.
\bibitem[{{European Commission}(1997)}]{europeancommission_directive_1997}
\bibinfo{author}{{European Commission}}, \bibinfo{year}{1997}.
\newblock \bibinfo{title}{Directive 96/92/{{EC}} concerning common rules for
  the internal market in electricity}.
\bibitem[{{European Commission}(2015)}]{europeancommission_commission_2015}
\bibinfo{author}{{European Commission}}, \bibinfo{year}{2015}.
\newblock \bibinfo{title}{Commission {{Regulation}} ({{EU}}) 2015/1222:
  {{Establishing}} a guideline on capacity allocation and congestion
  management}.
\bibitem[{{European Commission}(2019a)}]{europeancommission_commission_2019}
\bibinfo{author}{{European Commission}}, \bibinfo{year}{2019}a.
\newblock \bibinfo{title}{Commission {{Regulation}} ({{EU}}) 2019/943 on the
  internal market for electricity}.
\bibitem[{{European Commission}(2019b)}]{europeancommission_directive_2019}
\bibinfo{author}{{European Commission}}, \bibinfo{year}{2019}b.
\newblock \bibinfo{title}{Directive ({{EU}}) 2019/944 on common rules for the
  internal market for electricity.}
\bibitem[{Felice et~al.(2021)Felice, Peronato and Kavvadias}]{felice_jrc_2021}
\bibinfo{author}{Felice, M.D.}, \bibinfo{author}{Peronato, G.},
  \bibinfo{author}{Kavvadias, K.}, \bibinfo{year}{2021}.
\newblock \bibinfo{title}{{{JRC Hydro}}-Power Database}.
\newblock \bibinfo{type}{Dataset} \bibinfo{number}{Release v8}.
\bibitem[{Felten et~al.(2021)Felten, Osinski, Felling and
  Weber}]{felten_flowbased_2021}
\bibinfo{author}{Felten, B.}, \bibinfo{author}{Osinski, P.},
  \bibinfo{author}{Felling, T.}, \bibinfo{author}{Weber, C.},
  \bibinfo{year}{2021}.
\newblock \bibinfo{title}{The flow-based market coupling domain - {{Why}} we
  can't get it right}.
\newblock \bibinfo{journal}{Utilities Policy} \bibinfo{volume}{70},
  \bibinfo{pages}{101136}.
\bibitem[{{Gurobi Optimization LLC}(2018)}]{gurobioptimizationllc_gurobi_2018}
\bibinfo{author}{{Gurobi Optimization LLC}}, \bibinfo{year}{2018}.
\newblock \bibinfo{title}{Gurobi {{Optimizer Reference Manual}}}.
\bibitem[{Göke(2021)}]{goke_graphbased_2021}
\bibinfo{author}{Göke, L.}, \bibinfo{year}{2021}.
\newblock \bibinfo{title}{A graph-based formulation for modeling macro-energy
  systems}.
\newblock \bibinfo{journal}{Applied Energy} \bibinfo{volume}{301},
  \bibinfo{pages}{117377}.
\bibitem[{Göke et~al.(2021)Göke, Kendziorski, Kemfert and {von
  Hirschhausen}}]{goke_accounting_2021}
\bibinfo{author}{Göke, L.}, \bibinfo{author}{Kendziorski, M.},
  \bibinfo{author}{Kemfert, C.}, \bibinfo{author}{{von Hirschhausen}, C.},
  \bibinfo{year}{2021}.
\newblock \bibinfo{title}{Accounting for Spatiality of Renewables and Storage
  in Transmission Planning}.
\newblock \bibinfo{type}{{{arXiv}} Preprint} \bibinfo{number}{2108.04863}.
\bibitem[{Hainsch et~al.(2020)Hainsch, Brauers, Burandt, Göke, {von
  Hirschhausen}, Kemfert, Kendziorski, Löffler, Oei, Präger and
  Wealer}]{hainsch_make_2020}
\bibinfo{author}{Hainsch, K.}, \bibinfo{author}{Brauers, H.},
  \bibinfo{author}{Burandt, T.}, \bibinfo{author}{Göke, L.},
  \bibinfo{author}{{von Hirschhausen}, C.R.}, \bibinfo{author}{Kemfert, C.},
  \bibinfo{author}{Kendziorski, M.}, \bibinfo{author}{Löffler, K.},
  \bibinfo{author}{Oei, P.Y.}, \bibinfo{author}{Präger, F.},
  \bibinfo{author}{Wealer, B.}, \bibinfo{year}{2020}.
\newblock \bibinfo{title}{Make the {{European Green Deal}} Real: {{Combining}}
  Climate Neutrality and Economic Recovery}.
\newblock \bibinfo{type}{Politikberatung Kompakt} \bibinfo{number}{153}. {DIW
  Berlin}.
\bibitem[{Hofmann et~al.(2021)Hofmann, Hampp, Neumann, Brown and
  Hörsch}]{hofmann_atlite_2021}
\bibinfo{author}{Hofmann, F.}, \bibinfo{author}{Hampp, J.},
  \bibinfo{author}{Neumann, F.}, \bibinfo{author}{Brown, T.},
  \bibinfo{author}{Hörsch, J.}, \bibinfo{year}{2021}.
\newblock \bibinfo{title}{Atlite: {{A Lightweight Python Package}} for
  {{Calculating Renewable Power Potentials}} and {{Time Series}}}.
\newblock \bibinfo{journal}{Journal of Open Source Software}
  \bibinfo{volume}{6}, \bibinfo{pages}{3294}.
\bibitem[{Kunz et~al.(2017)Kunz, Kendziorski, Schill, Weibezahn, Zepter, {von
  Hirschhausen}, Hauser, Zech, Möst, Heidari, Felten and
  Weber}]{kunz_electricity_2017}
\bibinfo{author}{Kunz, F.}, \bibinfo{author}{Kendziorski, M.},
  \bibinfo{author}{Schill, W.P.}, \bibinfo{author}{Weibezahn, J.},
  \bibinfo{author}{Zepter, J.}, \bibinfo{author}{{von Hirschhausen}, C.},
  \bibinfo{author}{Hauser, P.}, \bibinfo{author}{Zech, M.},
  \bibinfo{author}{Möst, D.}, \bibinfo{author}{Heidari, S.},
  \bibinfo{author}{Felten, J.}, \bibinfo{author}{Weber, C.},
  \bibinfo{year}{2017}.
\newblock \bibinfo{title}{Electricity, {{Heat}} and {{Gas Sector Data}} for
  {{Modelling}} the {{German System}}}.
\newblock \bibinfo{type}{Data {{Documentation}}} \bibinfo{number}{92}. {DIW
  Berlin}.
\bibitem[{Lehner and Grill(2013)}]{lehner_global_2013}
\bibinfo{author}{Lehner, B.}, \bibinfo{author}{Grill, G.},
  \bibinfo{year}{2013}.
\newblock \bibinfo{title}{Global river hydrography and network routing:
  Baseline data and new approaches to study the world's large river systems}.
\newblock \bibinfo{journal}{Hydrological Processes} \bibinfo{volume}{27},
  \bibinfo{pages}{2171--2186}.
\bibitem[{Liu et~al.(2019)Liu, Andresen, Brown and
  Greiner}]{liu_highresolution_2019}
\bibinfo{author}{Liu, H.}, \bibinfo{author}{Andresen, G.B.},
  \bibinfo{author}{Brown, T.}, \bibinfo{author}{Greiner, M.},
  \bibinfo{year}{2019}.
\newblock \bibinfo{title}{A high-resolution hydro power time-series model for
  energy systems analysis: {{Validated}} with {{Chinese}} hydro reservoirs}.
\newblock \bibinfo{journal}{MethodsX} \bibinfo{volume}{6},
  \bibinfo{pages}{1370--1378}.
\bibitem[{Marjanovic et~al.(2018)Marjanovic, vom Stein, {van Bracht} and
  Moser}]{marjanovic_impact_2018}
\bibinfo{author}{Marjanovic, I.}, \bibinfo{author}{vom Stein, D.},
  \bibinfo{author}{{van Bracht}, N.}, \bibinfo{author}{Moser, A.},
  \bibinfo{year}{2018}.
\newblock \bibinfo{title}{Impact of an {{Enlargement}} of the {{Flow Based
  Region}} in {{Continental Europe}}}, in: \bibinfo{booktitle}{2018 15th
  {{International Conference}} on the {{European Energy Market}}}, pp.
  \bibinfo{pages}{1--5}.
\bibitem[{Matthes et~al.(2019)Matthes, Spieker, Klein and
  Rehtanz}]{matthes_impact_2019}
\bibinfo{author}{Matthes, B.}, \bibinfo{author}{Spieker, C.},
  \bibinfo{author}{Klein, D.}, \bibinfo{author}{Rehtanz, C.},
  \bibinfo{year}{2019}.
\newblock \bibinfo{title}{Impact of a {{Minimum Remaining Available Margin
  Adjustment}} in {{Flow}}-{{Based Market Coupling}}}, in:
  \bibinfo{booktitle}{2019 {{IEEE Milan PowerTech}}}, pp.
  \bibinfo{pages}{1--6}.
\bibitem[{{Rte} et~al.(2015){Rte}, {Amprion}, {Creos}, {Elia}, {TenneT} and
  {TransnetBW}}]{rte_cwe_2015}
\bibinfo{author}{{Rte}}, \bibinfo{author}{{Amprion}},
  \bibinfo{author}{{Creos}}, \bibinfo{author}{{Elia}},
  \bibinfo{author}{{TenneT}}, \bibinfo{author}{{TransnetBW}},
  \bibinfo{year}{2015}.
\newblock \bibinfo{title}{{{CWE Flow Based Market Coupling}} project:
  {{Parallel Run}} performance report}.
\bibitem[{Schönheit et~al.(2021)Schönheit, Kenis, Lorenz, Möst, Delarue and
  Bruninx}]{schonheit_fundamental_2021}
\bibinfo{author}{Schönheit, D.}, \bibinfo{author}{Kenis, M.},
  \bibinfo{author}{Lorenz, L.}, \bibinfo{author}{Möst, D.},
  \bibinfo{author}{Delarue, E.}, \bibinfo{author}{Bruninx, K.},
  \bibinfo{year}{2021}.
\newblock \bibinfo{title}{Toward a fundamental understanding of flow-based
  market coupling for cross-border electricity trading}.
\newblock \bibinfo{journal}{Advances in Applied Energy} \bibinfo{volume}{2},
  \bibinfo{pages}{100027}.
\bibitem[{Schönheit et~al.(2020)Schönheit, Weinhold and
  Dierstein}]{schonheit_impact_2020}
\bibinfo{author}{Schönheit, D.}, \bibinfo{author}{Weinhold, R.},
  \bibinfo{author}{Dierstein, C.}, \bibinfo{year}{2020}.
\newblock \bibinfo{title}{The impact of different strategies for generation
  shift keys ({{GSKs}}) on the flow-based market coupling domain: {{A}}
  model-based analysis of {{Central Western Europe}}}.
\newblock \bibinfo{journal}{Applied Energy} \bibinfo{volume}{258},
  \bibinfo{pages}{114067}.
\bibitem[{{Van den Bergh} et~al.(2016){Van den Bergh}, Boury and
  Delarue}]{vandenbergh_flowbased_2016}
\bibinfo{author}{{Van den Bergh}, K.}, \bibinfo{author}{Boury, J.},
  \bibinfo{author}{Delarue, E.}, \bibinfo{year}{2016}.
\newblock \bibinfo{title}{The {{Flow}}-{{Based Market Coupling}} in {{Central
  Western Europe}}: {{Concepts}} and definitions}.
\newblock \bibinfo{journal}{The Electricity Journal} \bibinfo{volume}{29},
  \bibinfo{pages}{24--29}.
\bibitem[{Weibezahn et~al.(2018)Weibezahn, Weinhold, Gerbaulet and
  Kunz}]{weibezahn_opsd_2018}
\bibinfo{author}{Weibezahn, J.}, \bibinfo{author}{Weinhold, R.},
  \bibinfo{author}{Gerbaulet, C.}, \bibinfo{author}{Kunz, F.},
  \bibinfo{year}{2018}.
\newblock \bibinfo{title}{{{OPSD}} Data Package: {{Conventional}} Power
  Plants}.
\newblock \bibinfo{type}{Dataset} \bibinfo{number}{Version: 2018-12-20}.
\bibitem[{Weinhold(2021)}]{weinhold_pomatodata_2021}
\bibinfo{author}{Weinhold, R.}, \bibinfo{year}{2021}.
\newblock \bibinfo{title}{{{PomatoData}} - {{GitHub Repository}}}.
\newblock \bibinfo{howpublished}{www.github.com/richard-weinhold/pomato\_data}.
\bibitem[{Weinhold and Mieth(2020a)}]{weinhold_fast_2020}
\bibinfo{author}{Weinhold, R.}, \bibinfo{author}{Mieth, R.},
  \bibinfo{year}{2020}a.
\newblock \bibinfo{title}{Fast {{Security}}-{{Constrained Optimal Power Flow
  Through Low}}-{{Impact}} and {{Redundancy Screening}}}.
\newblock \bibinfo{journal}{IEEE Transactions on Power Systems}
  \bibinfo{volume}{35}, \bibinfo{pages}{4574--4584}.
\bibitem[{Weinhold and Mieth(2020b)}]{weinhold_power_2020}
\bibinfo{author}{Weinhold, R.}, \bibinfo{author}{Mieth, R.},
  \bibinfo{year}{2020}b.
\newblock \bibinfo{title}{Power {{Market Tool}} ({{POMATO}}) for the
  {{Analysis}} of {{Zonal Electricity Markets}}}.
\newblock \bibinfo{type}{{{arXiv}} Preprint} \bibinfo{number}{2011.11594v1}.
\bibitem[{Weinhold and Mieth(2021)}]{weinhold_uncertaintyaware_2021}
\bibinfo{author}{Weinhold, R.}, \bibinfo{author}{Mieth, R.},
  \bibinfo{year}{2021}.
\newblock \bibinfo{title}{Uncertainty-{{Aware Capacity Allocation}} in
  {{Flow}}-{{Based Market Coupling}}}.
\newblock \bibinfo{type}{{{arXiv}} Preprint}.
\bibitem[{Wiegmans(2016)}]{wiegmans_gridkit_2016}
\bibinfo{author}{Wiegmans, B.}, \bibinfo{year}{2016}.
\newblock \bibinfo{title}{Gridkit {{Extract Of Entso}}-{{E Interactive Map}}}.
\newblock \bibinfo{type}{Dataset}. {Zenodo}.
\bibitem[{Wyrwoll et~al.(2018)Wyrwoll, Kollenda, Müller and
  Schnettler}]{wyrwoll_impact_2018}
\bibinfo{author}{Wyrwoll, L.}, \bibinfo{author}{Kollenda, K.},
  \bibinfo{author}{Müller, C.}, \bibinfo{author}{Schnettler, A.},
  \bibinfo{year}{2018}.
\newblock \bibinfo{title}{Impact of {{Flow}}-{{Based Market Coupling
  Parameters}} on {{European Electricity Markets}}}, in:
  \bibinfo{booktitle}{53rd {{International Universities Power Engineering
  Conference}}}, pp. \bibinfo{pages}{1--6}.

\end{thebibliography}

\end{document}